\documentclass[preprint,12pt]{elsarticle}

\usepackage{amssymb}

\usepackage{hyperref}
\newcommand\rf[1]{(\ref{eq:#1})}
\newcommand\lab[1]{\label{eq:#1}}
\newcommand\nonu{\nonumber}
\newcommand\br{\begin{eqnarray}}
\newcommand\er{\end{eqnarray}}
\newcommand\be{\begin{equation}}
\newcommand\ee{\end{equation}}

\newcommand\foot[1]{\footnotemark\footnotetext{#1}}
\newcommand\lb{\lbrack}
\newcommand\rb{\rbrack}

\newcommand\llb{\left\lbrack}
\newcommand\rrb{\right\rbrack}

\renewcommand\({\left(}
\renewcommand\){\right)}
\newcommand\bgv{\bigg\vert}              

\newcommand\bc{\begin{center}}
\newcommand\ec{\end{center}}

\renewcommand\a{\alpha}
\renewcommand\b{\beta}

\renewcommand\d{\delta}

\newcommand\vareps{\varepsilon}
\newcommand\g{\gamma}
\newcommand\G{\Gamma}

\newcommand\h{\frac{1}{2}}
\renewcommand\k{\kappa}
\renewcommand\l{\lambda}
\renewcommand\L{\Lambda}
\newcommand\m{\mu}
\newcommand\n{\nu}
\newcommand\om{\omega}

\newcommand\vp{\varphi}

\newcommand\pa{\partial}

\newcommand\pr{\prime}

\renewcommand\r{\rho}

\renewcommand\t{\tau}
\renewcommand\th{\theta}

\newcommand\wti{\widetilde}

\newcommand\cE{{\mathcal E}}

\newcommand\cJ{{\mathcal J}}

\newcommand{\ct}[1]{\cite{#1}}
\newcommand{\bib}[1]{\bibitem{#1}}

\newcommand\PRL[3]{\textsl{Phys. Rev. Lett.} \textbf{#1} (#2) #3}
\newcommand\NPB[3]{\textsl{Nucl. Phys.} \textbf{B#1} (#2) #3}

\newcommand\PRD[3]{\textsl{Phys. Rev.} \textbf{D#1} (#2) #3}

\newcommand\PLB[3]{\textsl{Phys. Lett.} \textbf{#1B} (#2) #3}
\newcommand\CQG[3]{\textsl{Class. Quantum Grav.} \textbf{#1} (#2) #3}

\newcommand\AoP[3]{\textsl{Ann. of Phys.} \textbf{#1} (#2) #3}

\newcommand\IJMPA[3]{\textsl{Int. J. Mod. Phys.} \textbf{A#1} (#2) #3}

\newcommand\MPLA[3]{\textsl{Mod. Phys. Lett.} \textbf{A#1} (#2) #3}

\newcommand\xdot{\stackrel{.}{x}}
\newcommand\ydot{\stackrel{.}{y}}

\newcommand\udot{\stackrel{.}{u}}
\newcommand\vdot{\stackrel{.}{v}}

\newcommand\xdotdot{\stackrel{..}{x}}
\newcommand\ydotdot{\stackrel{..}{y}}

\begin{document}

\begin{frontmatter}



\title{Charge-Confining Gravitational Electrovacuum\\ Shock Wave}


\author[BGU]{Eduardo Guendelman\corref{cor1}}
\ead{guendel@bgu.ac.il}
\ead[url]{http://eduardo.guendelman.com}
\cortext[cor1]{Corresponding author -- tel. +972-8-647-2508, fax +972-8-647-2904.}
\address[BGU]{Department of Physics, Ben-Gurion University of the Negev,
P.O.Box 653, IL-84105 ~Beer-Sheva, Israel}

\author[INRNE]{Emil Nissimov}
\ead{nissimov@inrne.bas.bg}
\ead[url]{http://theo.inrne.bas.bg/~nissimov/}
\author[INRNE]{Svetlana Pacheva}
\ead{svetlana@inrne.bas.bg}
\ead[url]{http://theo.inrne.bas.bg/~svetlana/}
\address[INRNE]{Institute for Nuclear Research and Nuclear Energy, Bulgarian Academy
of Sciences, Boul. Tsarigradsko Chausee 72, BG-1784 ~Sofia, Bulgaria}

\begin{abstract}
In previous publications we have extensively studied spherically symmetric
solutions of gravity coupled to a non-standard type of non-linear
electrodynamics containing a square root of the ordinary Maxwell Lagrangian
(the latter is known to yield QCD-like confinement in flat space-time).
A class of these solutions describe non-standard black holes of 
Reissner-Nordstr{\"o}m-\-(anti-)\-de-Sitter type with an additional constant
radial vacuum electric field, in particular, a non-asymptotically flat 
Reissner-Nordstr{\"o}m-type black hole. Here we study the ultra-relativistic
boost (Aichelburg-Sexl-type) limit of the latter and show that, unlike the
ordinary Reissner-Nordstr{\"o}m case, we obtain a {\em gravitational
electrovacuum shock wave} as a result of the persistence of the gauge field 
due to the ``square-root'' Maxwell Lagrangian term. Next, we show that this
gravitational electrovacuum shock wave {\em confines} charged test particles
(both massive and massless) within a finite distance from its front.
\end{abstract}

\begin{keyword}
\PACS 11.25.-w \sep 04.70.-s \sep 04.50.+h
\end{keyword}

\end{frontmatter}


\section{Introduction}
\label{intro}



In his analysis in Ref.\ct{tHooft} `t Hooft has shown that in any effective 
quantum gauge theory, which is able to describe linear confinement phenomena, 
the energy density of electrostatic field configurations should be a linear function
of the electric displacement field $\vec{D}$ in the infrared region. 
In particular, a consistent quantum approach has been developed in \ct{tHooft},
where the electric displacement field appears as an ``infrared counterterm''.

The simplest way to realize these ideas in flat space-time was proposed in
Refs.\ct{GG} by considering a nonlinear gauge theory with an action:
\br
S = \int d^4 x\, L(F^2) \quad ,\quad
L(F^2) = -\frac{1}{4} F^2 - \frac{f_0}{2} \sqrt{-F^2} \; ,
\lab{GG-flat} \\
F^2 \equiv F_{\m\n} F^{\m\n} \quad ,\quad 
F_{\m\n} = \pa_\m A_\n - \pa_\n A_\m  \; ,
\nonu
\er
where $f_0$ is a positive coupling constant. 
It has been shown in the first reference in \ct{GG} that the ``square root''
Maxwell term $\sqrt{-F^2}$ naturally arises as a result of spontaneous
breakdown of scale symmetry of the original scale-invariant Maxwell action. 
For static field configurations the model \rf{GG-flat} yields an energy density
that indeed contains a term linear w.r.t. $|\vec{D}|$, where
$\vec{D} = \vec{E} - \frac{f_0}{\sqrt{2}}\frac{\vec{E}}{|\vec{E}|}$. 
The model \rf{GG-flat} produces a confining effective potential containing a 
Coulomb plus a linear one, which is of the form of the well-known
``Cornell'' potential \ct{cornell-potential} in 
the phenomenological description of quarkonium systems (see second reference in 
\ct{GG}). Let us remark that one could start with the non-Abelian version of 
the action \rf{GG-flat} as well, since the non-Abelian theory can effectively be
reduced to an Abelian one in the static spherically symmetric case as pointed out 
in the second reference in \ct{GG}.

Coupling of the nonlinear gauge field system \rf{GG-flat} to ordinary Einstein 
gravity as well as to $f(R) = R + \a R^2$ gravity was recently studied in
\ct{grav-cornell,hiding-hide-confine,f(R)-maxwell} 
\foot{Let us also mention the recent papers \ct{halilsoy} where coupling of ordinary 
Einstein gravity to the  pure ``square-root'' gauge field Lagrangian is discussed.
One of the new interesting features in this model is the existence of dyonic 
solutions.}, where the following interesting new features of the pertinent static 
spherically symmetric solutions have been found:

(i) Appearance of a constant radial vacuum electric field (in addition to the
Coulomb one) in charged black holes within Reissner-Nordstr{\"o}m-\-(anti-)\-de-Sitter
space-times, in particular, in electrically neutral black holes with 
Schwarzschild-(anti-)de-Sitter geometry. Let us particularly stress, that 
constant radial electric fields {\em do not} exist as solutions of ordinary
Maxwell electrodynamics.

(ii) Novel mechanism of {\em dynamical generation} of cosmological constant
through the nonlinear gauge field dynamics of the ``square-root'' gauge field
term. 

(iii) Even in case of vanishing effective cosmological constant the resulting 
Reissner-Nordstr{\"o}m-type black hole, apart from carrying an additional
constant radial vacuum electric field, turns out to be {\em non-asymptotically flat}
-- a feature resembling the gravitational effect of a hedgehog \ct{hedgehog}. 

(iv) New ``tube-like universe'' solutions of Levi-Civita-Bertotti-Robinson 
\ct{LC-BR} type.

(v) Coupling the gravity/non-linear gauge field system self-consistently to 
{\em lightlike} branes\foot{For a systematic reparametrization-invariant
world-volume Lagrangian formulation of {\em lightlike} branes, see 
Refs.\ct{LL-main}.} produces charge-hiding" and
charge-confining ``thin-shell'' wormhole solutions displaying 
QCD-like charge confinement \ct{hiding-hide-confine}. Similar effect is
obtained also via coupling to ordinary Nambu-Goto branes \ct{mahary-hiding}.

Coupling of the nonlinear gauge field system \rf{GG-flat} to 
$f(R) = R + \a R^2$ gravity plus a dilaton \ct{f(R)-maxwell} produces in addition 
to the above properties (i)-(v) also an appearance of dynamical effective gauge 
couplings as well as confinement-deconfinement transition effect as functions of 
the dilaton vacuum expectation value. Finally, non-singular wormhole solutions 
that do not require exotic matter have been found in a scalar curvature square 
plus Ricci squared model \ct{ricci-square}.

One should notice that all of the above solutions are both static and 
spherically symmetric. The purpose of the present note is to study time-dependent 
``shock wave'' type solutions in the gravity/non-linear gauge field system.

Gravitational shock waves are of particular relevance (starting with the classic 
paper \ct{aichel-sexl}) in modern field and string theory, especially due to their
role in the description of impulsive lightlike signals in general relativity 
\ct{barrabes-hogan}, high-energy scattering at Planck energies and 
ultra-relativistic heavy ion collisions (see \textsl{e.g.} \ct{ultra-planck-RHIS} 
and references therein), {\em etc}. 

Gravitational shock wave solutions can be obtained from static spherically 
symmetric ones by an ultra-relativistic boost procedure originating from the 
paper by Aichelburg and Sexl \ct{aichel-sexl} (boost of ordinary Schwarzschild 
metric, see also \ct{dray-thooft}), which was subsequently
generalized to ultra-relativistic boosts of Reissner-Nordstr{\"o}m, 
Reissner-Nordstr{\"o}m-(anti-)de-Sitter and Kerr-Newman geometries
\ct{lousto-sanchez,dS-boost,kerr-boost}.

When performing an ultra-relativistic boost of the ordinary Reissner-Nordstr{\"o}m
metric, in order to get a well defined limit one must appropriately rescale 
the mass ($m \sim 1/\g$) and the charge ($Q^2 \sim 1/\g$) with the
Lorentz-boost factor $\g = (1-w^2)^{-\h} \to \infty$ when $w \to 1$.
This implies the vanishing of the electromagnetic field in the
ultra-relativistic limit, although a finite $\d$-function contribution of the latter 
remains in the energy-momentum tensor, so that it contributes only
to the shape of the gravitational shock wave profile.

The first task in the present note is to study the ultra-relativistic boost
limit of the simplest non-trivial new solution obtained by coupling gravity to
the non-linear gauge field system \rf{GG-flat}, namely, the above mentioned
non-asymptotically flat Reissner-Nordstr{\"o}m type black hole carrying an additional
constant radial vacuum electric field apart from the Coulomb one.
Unlike ordinary Reissner-Nordstr{\"o}m case where the electromagnetic field
vanishes in the boost limit, here the resulting shock wave we obtain is a
non-trivial gravitational electrovacuum one with a non-zero electric and
magnetic fields originating from the constant radial vacuum electric field
in the rest-frame.

The second main task is to study the dynamics of charged test particles
(both massive and massless) in the newly obtained gravitational electrovacuum
shock wave background. We show that the latter exhibits a QCD-like charge-confining
feature.

\section{Ultra-Relativistic Limit of Non-standard Black Hole with Confining
Electric Potential}
\label{UR-limit}

The simplest coupling to gravity of the nonlinear gauge field system with a square 
root of the Maxwell term \rf{GG-flat} known to produce QCD-like confinement
in flat space-time \ct{GG} is given by the action 
(we use units with Newton constant $G_N=1$):
\br
S = \int d^4 x \sqrt{-g} \Bigl\lb \frac{R(g) - 2\L_0}{16\pi} + L(F^2)\Bigr\rb 
\quad ,\quad
L(F^2) = - \frac{1}{4} F^2 - \frac{f_0}{2} \sqrt{\vareps F^2} \; ,
\lab{gravity+GG} \\
F^2 \equiv F_{\k\l} F_{\m\n} g^{\k\m} g^{\l\n} \quad ,\quad 
F_{\m\n} = \pa_\m A_\n - \pa_\n A_\m \; .
\nonu
\er
Here $R(g)=R_{\m\n} g^{\m\n}$ is the scalar curvature of the space-time metric
$g_{\m\n}$, $R_{\m\n}$ is the Ricci tensor and $g \equiv \det\Vert g_{\m\n}\Vert$;
the sign factor $\vareps = \pm 1$
in the square-root term in \rf{gravity+GG} corresponds to ``magnetic'' or ``electric''
dominance -- in what follows we consider the  $\vareps = -1$ case; $f_0$ is a 
positive coupling constant measuring the strength of charge confinement. 

Let us stress that we {\em do not} need to introduce {\em any} bare 
cosmological constant $\L_0$ in \rf{gravity+GG} since the ``square-root'' Maxwell term 
dynamically generates a {\em non-zero effective cosmological constant} 
$\L_{\rm{eff}} = 2\pi f_0^2$ \ct{hiding-hide-confine}. 
The role of the bare $\L_0$ is just shifting 
the effective $\L_{\rm{eff}}$ (see Eqs.\rf{CC-eff} below).


\vspace{.1in}
The equations of motion corresponding to \rf{gravity+GG} read:
\be
R_{\m\n} - \h g_{\m\n} R + \L_0\, g_{\m\n} = 8\pi T_{\m\n} \; ,
\lab{einstein-eqs}
\ee
\be
T_{\m\n} = \Bigl( 1 - \frac{f_0}{\sqrt{-F^2}}\Bigr) F_{\m\k} F_{\n\l} g^{\k\l}
- \frac{1}{4} \Bigl( F^2 + 2f_0 \sqrt{-F^2}\Bigr) g_{\m\n} \; ,
\lab{stress-tensor-F}
\ee
and
\be
\pa_\n\(\sqrt{-g}\Bigl(1-\frac{f_0}{\sqrt{-F^2}}\Bigr) F_{\k\l}g^{\m\k}g^{\n\l}\)=0
\; .
\lab{GG-eqs}
\ee

In our preceding papers \ct{grav-cornell,hiding-hide-confine} we have shown that 
the gravity/nonlinear-gauge-field system \rf{gravity+GG} possesses static 
spherically symmetric solutions with a radial electric field containing {\em both}
Coulomb and {\em constant} ``vacuum'' pieces:
\be
F_{0r} = \frac{\vareps_F f_0}{\sqrt{2}} - \frac{Q}{\sqrt{4\pi}\, r^2} 
\quad ,\quad \vareps_F \equiv \mathrm{sign}(F_{0r}) = - \mathrm{sign}(Q) \; ,
\lab{cornell-sol}
\ee
and the space-time metric:
\br
ds^2 = - A(r) dt^2 + \frac{dr^2}{A(r)} + r^2 \bigl(d\th^2 + \sin^2 \th d\vp^2\bigr)
\; ,
\lab{spherical-static} \\
A(r) = 1 - \sqrt{8\pi}|Q|f_0 - \frac{2m}{r} + \frac{Q^2}{r^2} 
- \frac{\L_{\mathrm{eff}}}{3} r^2 
\quad ,\quad \L_{\mathrm{eff}} = \L_0 + 2\pi f_0^2 \; ,
\lab{CC-eff}
\er
is that of a Reissner-Nordstr{\"o}m-(anti-)de-Sitter-type 
black hole depending on the sign of the {\em dynamically generated}
effective cosmological constant $\L_{\mathrm{eff}}$.

\begin{itemize}
\item
Solution \rf{cornell-sol}--\rf{CC-eff} represents a {\em non-standard} black
hole whose ``vacuum'' electric field persists even in the electrically
neutral case ($Q=0$ in \rf{cornell-sol} and \rf{CC-eff}).
\item
In the presence of the bare cosmological constant $\L_0$ in the action \rf{gravity+GG} 
its only effect is shifting of the dynamically generated cosmological constant 
$2\pi f_0^2$ in second Eq.\rf{CC-eff}.
\item
Solution \rf{cornell-sol}--\rf{CC-eff} becomes a non-standard
Reissner-Nordstr{\"o}m-type black hole with additional constant vacuum radial 
electric field in spite of the presence of {\em negative} bare cosmological constant
$\L_0 < 0$ with $|\L_0| = 2\pi f_0^2$, \textsl{i.e.}, $\L_{\mathrm{eff}}=0$ in
\rf{CC-eff}. Another noteworthy feature is its {\em non-flat} space-time asymptotics
due to the ``leading'' constant term in the metric coefficient \rf{CC-eff} being 
different from 1 when $Q \neq 0$. This resembles the effect on gravity 
produced by a spherically symmetric ``hedgehog'' configuration of a nonlinear 
sigma-model scalar field with $SO(3)$ symmetry 
(see Refs.\ct{Ed-Rab-hedge-barriola-vilenkin}).
\end{itemize}

In what follows we will consider in more detail the latter special case 
--  the non-asymptotically flat Reissner-Nordstr{\"o}m-type black hole with
additional constant vacuum radial electric field ($\L_0 < 0$ with 
$|\L_0| = 2\pi f_0^2$ in \rf{gravity+GG}, \textsl{i.e.}, $\L_{\mathrm{eff}}=0$ in 
\rf{CC-eff}).

To derive the ultra-relativistic boost limit of \rf{cornell-sol}--\rf{CC-eff}
we follow the approach of Aichelburg-Sexl \ct{aichel-sexl} and Lousto-Sanchez
\ct{lousto-sanchez}. The first step is to transform to isotropic-like
coordinates $(t,r) \to (t^{\pr}, r^{\pr})$:
\br
t = \frac{1}{\sqrt{c_0}}\, t^{\pr\,\sqrt{c_0}} \quad ,\quad
c_0 \equiv 1 - \sqrt{8\pi}|Q|f_0 \; ,
\lab{t-prime} \\
r = \sqrt{c_0}\, r^{\pr\,\sqrt{c_0}}\, \Bigl\lb 1 +
\frac{m}{2 c_0^{3/2} r^{\pr\,\sqrt{c_0}}}\Bigr\rb^2 
- \frac{Q^2}{4 c_0^{3/2} r^{\pr\,\sqrt{c_0}}} \; , 
\lab{r-prime} \\
x^\pr = r^\pr \cos \vp \sin \th \; ,\; y^\pr = r^\pr \sin \vp \sin \th \; ,\;
z^\pr =  r^\pr \cos \th \; ,
\nonu
\er
so that \rf{cornell-sol}--\rf{CC-eff} becomes:
\br
ds^2 = - {\wti A}(t^\pr,r^\pr) {dt^\pr}^2 + {\wti B}(r^\pr)
\( {dx^\pr}^2 + {dy^\pr}^2 + {dz^\pr}^2 \) \; ,
\lab{iso-static} \\
{\wti A}(t^\pr,r^\pr) = t^{\pr\,2(\sqrt{c_0}-1)} A\bigl( r(r^\pr)\bigr)
\quad ,\quad  {\wti B}(r^\pr) = \frac{ r^2(r^\pr)}{r^{\pr\,2}}  \; ,
\lab{iso-static-def} 
\er
\br
F = F_{0r} dt \wedge dr =
\Bigr\lb\frac{\vareps_F f_0}{\sqrt{2}} - \frac{Q}{\sqrt{4\pi}\, r^2(r^\pr)}\Bigr\rb
\({\wti A}{\wti B}\)^{1/2}
\frac{1}{r^\pr}\bigl\lb dt^\pr\wedge (x^\pr dx^\pr + y^\pr dy^\pr 
+ z^\pr dz^\pr)\bigr\rb
\nonu \\
{}
\lab{iso-cornell-sol}
\er

The next step is to perform a Lorentz boost in \textsl{e.g.}
$x^\pr$-direction:
\be
(t^\pr,x^\pr) \to ({\wti t},{\wti x}) \quad ,\quad
t^\pr = \g ({\wti t} - w {\wti x}) \; , \;
x^\pr = \g ({\wti x} - w {\wti t}) \; , \; 
\g \equiv (1 - w^2)^{-\h} \; ,
\lab{boost}
\ee
and then take the ultra-relativistic limit $w \to 1$,
\textsl{i.e.}, $\g \to \infty$, by rescaling the black hole parameters
$m = p/\g\, ,\, Q^2 = q^2/\g$ (with $p,q$ -- finite constants) and
using the distributional limits: 
\br
\frac{\g}{r^{\pr\,s}} \to 
\d (u) \r^{1-s}\, \frac{\G(\h) \G(\frac{s-1}{2})}{\G(\frac{s}{2})} 
\;\;\; {\rm for}\; s>1 \quad ,\quad
\frac{\g}{r^\pr} \to \frac{1}{|u|} - \d (u) \ln \r^2 \; .
\lab{distrib-eqs}
\er
Here the following notations are introduced:
\br
v = {\wti t} + {\wti x} \; ,\; u = {\wti t} - {\wti x} \quad ;\quad 
\r \equiv \sqrt{y_1^2 + y_2^2} \; ,\; (y_i)_{i=1,2} \equiv 
{\underline y} \equiv (y^\pr,z^\pr) \; ,
\lab{lc-def}\\
r^\pr = \sqrt{x^{\pr\, 2}+\r^2} \approx \sqrt{\g^2 u^2 + \r^2} \;\;
{\rm for ~large} \; \g \;.
\phantom{aaaaaaaaaaaa}
\nonu
\er

The choice of the $t$-coordinate transformation \rf{t-prime} (which does not
have an analog in the ordinary Reissner-Nordstr{\"o}m case) has been made in 
such a way as to obtain a well-defined ultra-relativistic limit for the
metric \rf{iso-static}-\rf{iso-static-def}. This limit turns out to coincide
with that of the ordinary Reissner-Nordstr{\"o}m metric \ct{lousto-sanchez}:
\be
ds^2 = - dv du - h(\r)\d (u) du^2 + {dy_1}^2 + {dy_2}^2 \quad ,\quad
h(\r) = 8p \ln \r + \frac{3\pi q^2}{2\r}
\lab{UR-RN-metric}
\ee
with non-zero Christoffel symbols:
\be
\G^v_{uu} = h(\r) \d^{\pr}(u) \quad ,\quad \G^v_{ui}=\pa_i h\,\d(u)
\quad ,\quad \G^i_{uu} = \h \pa_i h\,\d(u) \; .
\lab{christoffel}
\ee
As in the standard case \ct{aichel-sexl,lousto-sanchez}, in order to remove
the $\frac{1}{|u|}$ term in the $g_{uu}$ metric component coming from second relation
\rf{distrib-eqs}, one has to perform an additional coordinate transformation in
the $(v,u)$ subspace: $ v \to v + 4p \ln |u|\; ,\; u \to u$.

Concerning the electromagnetic field, the Coulomb piece in \rf{iso-cornell-sol} 
vanishes in the ultra-relativistic limit as it happens in the ordinary 
Reissner-Nordstr{\"o}m case \ct{lousto-sanchez}, however now we are left with a 
non-trivial limit:
\be
F_{vu} = \frac{\vareps_F f_0}{2\sqrt{2}}\, {\rm sign}(u) \;\; ,\;\;
F_{ui} = y_i \frac{\vareps_F f_0}{\sqrt{2}} \Bigl(\frac{1}{|u|} - \d (u) \ln \r^2\Bigr)
\; .
\lab{UR-EM-field}
\ee
The electromagnetic field potential corresponding to \rf{UR-EM-field} is:
\br
F_{vu} = - \pa_u A_v \quad ,\quad A_v = -\frac{\vareps_F f_0}{2\sqrt{2}}\,|u| \; ,
\lab{UR-EM-v} \\
F_{ui} = - \pa_i A_u \quad ,\quad A_u = \frac{\vareps_F f_0}{2\sqrt{2}}\,
\r^2 \Bigl\lb - \frac{1}{|u|} + \d (u)\, (\ln \r^2 - 1)\Bigr\rb \; .
\lab{UR-EM-i}
\er
In terms of ordinary definitions of electric and magnetic fields
${\vec E}=\( E_{\wti x}, E_{y^\pr}, E_{z^\pr}\)$,
${\vec B}=\( B_{\wti x}, B_{y^\pr}, B_{z^\pr}\)$ we have:
\br
E_{\wti x} = \frac{\vareps_F f_0}{\sqrt{2}}\; {\rm sign}({\wti t}-{\wti x})
\quad ,\quad B_{\wti x} = 0 \; ,
\nonu \\
E_{y^\pr} = 
\frac{\vareps_F f_0}{\sqrt{2}}\, y^\pr \Bigl\lb \ln \r^2\,\d ({\wti t}-{\wti x})
- \frac{1}{|{\wti t}-{\wti x}|}\Bigr\rb = B_{z^\pr} \; ,
\lab{E-B-shock} \\
E_{z^\pr} = 
\frac{\vareps_F f_0}{\sqrt{2}}\, z^\pr \Bigl\lb \ln \r^2\,\d ({\wti t}-{\wti x})
- \frac{1}{|{\wti t}-{\wti x}|}\Bigr\rb = - B_{y^\pr} \; ,
\nonu
\er
Let us stress that the ultra-relativistic boost limit of $F_{\m\n}$ obtained
in \rf{UR-EM-field} (or \rf{E-B-shock}) is a particular {\em vacuum} solution 
$F^2 = - f_0^2$ of the nonlinear gauge field Eqs.\rf{GG-eqs}, which {\em does not
exist} as a solution in ordinary Maxwell electrodynamics.
According to Eq.\rf{stress-tensor-F} its contibution to the total energy-momentum
tensor is $-\frac{1}{4}f_0^2 g_{\m\n}$ which exactly cancells the contribution of the
``bare'' cosmological constant term in Einstein Eqs.\rf{einstein-eqs}
we started with ($\L_0 = - 2\pi f_0^2$).

Thus, the ultra-relativistic boost of the non-asymptotically flat
Reissner-Nordstr{\"o}m-type black hole with an additional constant ``vacuum''
radial electric field \rf{cornell-sol}--\rf{CC-eff} describes a {\em gravitational 
electrovacuum shock wave}~ given by \rf{UR-RN-metric}--\rf{UR-EM-field}.

\section{Charged Test-Particle Dynamics in the Electrovacuum Gravitational Shock Wave}
\label{test-particle}

We now consider the motion of charged test particles in the background of
the gravitational electrovacuum shock wave \rf{UR-RN-metric}--\rf{UR-EM-i}. 
Here we study the case of test particle charge $q_0$ opposite to the
pre-boost Reissner-Nordstr{\"o}m charge $Q$, \textsl{i.e.}:
\be
{\rm sign}(q_0) = \vareps_F = - {\rm sign}(Q)
\lab{test-charge-plus}
\ee
according to \rf{cornell-sol}  
\foot{Note that the sign factor $\vareps_F$ in Eqs.\rf{UR-EM-field}-\rf{E-B-shock}
is the only legacy of the Reissner-Nordstr{\"o}m charge $Q$ in the boosted
electromagnetic field after performing the ultra-relativistic boost limit.}.
For definiteness we take ${\rm sign}(q_0) = \vareps_F = - {\rm sign}(Q)=1$.

The pertinent point-particle action reads:
\br
S_{\rm particle} = 
\int d\l \Bigl\lb \frac{1}{2e} g_{\m\n} \frac{dx^\m}{d\l} \frac{dx^\n}{d\l}
- \h e m_0^2 + q_0 \frac{dx^\m}{d\l} A_\m \Bigr\rb
\nonu \\
=  \int d\l \Bigl\lb \frac{1}{2e}\Bigl( - \frac{dv}{d\l} \frac{du}{d\l}
- h(\r)\d(u) \(\frac{du}{d\l}\)^2 + \frac{dy_i}{d\l} \frac{dy_i}{d\l}\Bigr)\Bigr.
\nonu \\
\Bigl. - \h e m_0^2 + q_0 \frac{dv}{d\l} A_v (u) + 
\frac{du}{d\l} q_0 A_u (u,\r) \Bigr\rb \; .
\lab{particle-action}
\er
Here $\l$ denotes the world-line parameter, 
$e$ is the world-line ``einbein'', $m_0$ is the test-particle mass, in particular,
$m_0$ could be zero, and $\r, h(\r)\, ,A_v\, ,A_u$ are given in \rf {lc-def}, 
\rf{UR-RN-metric}, \rf{UR-EM-v}, \rf{UR-EM-i} respectively.

Upon introducing world-line proper-time-like parameter $\t$ with 
$\frac{d\t}{d\l} = e$ \foot{This allows simultaneous treatment of both the massive
and massless case; the genuine proper-time parameter in the massive case is ${\bar\t}$
with $\frac{d{\bar\t}}{d\l} = em_0$.} the equations of motion w.r.t. $x^\m$ (geodesic)
and $e$ (mass-shell constraint) read:
\br
\xdotdot^\m + \G^\m_{\n\l}\xdot^\n \xdot^\l 
- q_0 \xdot^\n F_{\l\n}g^{\l\m}=0 \; ,
\lab{x-eqs} \\
g_{\m\n} \xdot^\m \xdot^\n + m_0^2 = 0 \;\; \longrightarrow \;\;
-\vdot\udot - h(\r) \d (u) \udot^2 + {\underline \ydot}^2 + m_0^2 = 0 \; ,
\lab{massive-eq}
\er
where the dots represent $\frac{d}{d\t}$, and $\G^\m_{\n\l}$ and $F_{\l\n}$
are given in \rf{christoffel}-\rf{UR-EM-field}.


We have two conservation laws due to invariance of \rf{particle-action} w.r.t.
$v$-translation (``light-cone energy'' $\cE$ conservation) and w.r.t. rotation in the
$(y_1,y_2)$-subspace (angular momentum $\cJ$ conservation), which according to Noether
theorem read:
\be
\cE = \h \udot - q_0 A_v \quad ,\quad \cJ =
y_1\,{\stackrel{.}{y_2}} -  y_2\,{\stackrel{.}{y_1}} 
\equiv {\underline y}\times {\stackrel{.}{\underline y}} \; ,
\lab{Noether}
\ee
(here $\times$ indicates two-dimensional vector product).
The first ``energy''-conservation Eq.\rf{Noether} upon substituting the
expression for $A_v$ from \rf{UR-EM-v} reads explicitly:
\be
\udot = - a\,|u| + 2\cE \quad ,\quad a \equiv \frac{q_0 f_0}{\sqrt{2}}
\lab{u-eq}
\ee
(henceforth we will use systematically the short-hand notation $a$ for the 
constant defined in \rf{u-eq}).

The ``mass-shell'' constraint Eq.\rf{massive-eq} yields a first-order
equation for $v(\t)$:
\be
\vdot = - h(\r)\udot\d(u) + \frac{1}{\udot}\,\Bigl({\underline \ydot}^2 +m_0^2\Bigr) \; ,
\lab{v-eq}
\ee
where $y_i (\t)$ is a solution of Eqs.\rf{x-eqs} for $\m=i$ (using expression
\rf{UR-EM-i} for $F_{ui}$):
\be
\ydotdot_i + a \frac{\udot}{|u|} y_i + \d(u)
\Bigl\lb \h \pa_i h (\r)\,\udot^2 - a \udot y_i \ln\r^2\Bigr\rb  = 0 \; .
\lab{massive-eqs-i}
\ee

The form of the test particle trajectory depends significantly on the
initial condition for $u(\t)$: $u_0 = u(0)$. 
There are three types of initial conditions:
\begin{itemize}
\item
(i) $|u_0| < 2\cE/a$ -- this is the case which will be treated in some
detail below. In this case the particle trajectory  ``pierces'' the shock 
wave, \textsl{i.e.},  $u(\t_0) = 0$ at some value $\t_0$ of the proper time 
(world-line) parameter. We find 
$|u(\t)| < 2\cE/a$ for all $\t$ (see Eq.\rf{u-confined} below) implying 
{\em confinement (trapping)} of the charged particle (both massive and massless)
within finite distance $2\cE/a$ from the shock wave.
\item
(ii) $|u_0| = 2\cE/a$ -- this initial condition by consistency applies only
for massless charged test particles. Now we have $|u(\t)| = 2\cE/a = {\rm const}$ and
$y_i (\t) = {\rm const}$ for all $\t$, $v(\t)$ is an arbitrary linear function of $\t$,
\textsl{i.e.}, this is free lightlike motion parallel to the shock wave at a fixed
``critical'' distance $2\cE/a$ from the latter.
\item
(iii) $|u_0| > 2\cE/a$ -- in this case $|u(\t)| > 2\cE/a$ for all $\t$ 
(see {\em Appendix}),\textsl{i.e.}, the particle trajectory never 
``pierces'' the shock wave and, since now the proper time (world-line) parameter $\t$
flows opposite the ``laboratory'' time $t$, this motion in fact corresponds to a 
repulsion of an anti-particle (cf. Ref.\ct{QED-feynman}).
\end{itemize}

We start with case (i). With the initial condition
$|u(0)| \equiv |u_0| < 2\cE/a$ the solution of Eq.\rf{u-eq} reads:
\be
u(\t) = {\rm sign}(\t-\t_0) \frac{2\cE}{a} \( 1 - e^{-a|\t-\t_0|}\) \; ,
\lab{u-sol}
\ee
where:
\be
u(\t_0) = 0 \quad ,\quad \t_0 = {\rm sign}(u_0) \frac{1}{a}
\ln\Bigl(\frac{2\cE/a - |u_0|}{2\cE/a}\Bigr) \; .
\lab{u-sol-0}
\ee
Eq.\rf{u-sol} implies:
\be
|u(\t)| = \frac{2\cE}{a} \( 1 - e^{-a|\t-\t_0|}\) \leq \frac{2\cE}{a} \;\;
{\rm for ~all}\; \t \in (-\infty,+\infty) \; .
\lab{u-confined}
\ee
Henceforth, for simplicity we take $\t_0 = 0$, \textsl{i.e.}, $u_0 = 0$.

Eq.\rf{v-eq} yields for $v(\t)$:
\be
v(\t) = - h(\r_0) \th (\t) + 
\frac{1}{2\cE} \int d\t^\pr e^{a|\t^\pr|} \({\underline \ydot}^2 + m_0^2\) 
\quad ,\quad \r_0 = |{\underline y}(0)| \; ,
\lab{v-integr}
\ee
whereas Eqs.\rf{massive-eqs-i} for $y_i$  acquire the form (taking into account 
\rf{u-sol} and using $\d (u) = |\udot|^{-1} \d (\t)$):
\be
{\underline \ydotdot} + \om^2 (\t) {\underline y}(\t) + \d (\t)\,{\underline y}(0)
\Bigl\lb \frac{\cE}{\r_0} \frac{d h}{d\r}\bgv_{\r=\r_0} - 2a\ln\r_0 \Bigr\rb = 0
\quad , \;\; 
\om^2 (\t) \equiv \frac{a^2}{e^{a|\t|}-1} \; .
\lab{y-eqs}
\ee

Asymptotically for large $|\t|$, $y_i (\t)$ is linear function:
\be
{\underline y}(\t) \approx {\underline \a}_{(\pm)} \t + {\underline \b}_{(\pm)}
\;\; {\rm for}\;\; \t \to \pm \infty \;\; ,\;\; 
{\underline \a}_{(\pm)}, {\underline \b}_{(\pm)} = {\rm const} \; .
\lab{y-large-tau}
\ee
In the vicinity of $\t=0$ the solution of \rf{y-eqs} is given in terms of Bessel
functions (since $\om^2 (\t) \approx a/|\t|$ for $\t \to 0$):
\be
{\underline y}(\t) = \sqrt{|\t|}\Bigl\lb \frac{1}{\sqrt{a}}
{\underline c}_{(\pm)} J_1 (2\sqrt{a|\t|}) + \sqrt{a}\,
{\underline d}_{(\pm)} Y_1 (2\sqrt{a|\t|})\Bigr\rb \; ,
\lab{y-small-tau}
\ee
where ${\underline c}_{(\pm)}, {\underline d}_{(\pm)} = {\rm const}$ and the
subscripts $(\pm)$ correspond to $\t>0$ and $\t<0$, respectively.

Using the series expansions of the Bessel functions and requiring continuity
of ${\underline y}(\t)$ at $\t=0$, which implies
${\underline y}_0 \equiv {\underline y}(0) = - \frac{1}{\pi} {\underline d}_{(\pm)}$,
we have for small $\t$ ($\g_0$ denotes the Euler constant):
\br
{\underline y}(\t) \approx {\underline y}_0 +
\t \Bigl\lb \th (\t){\underline c}_{(+)} - \th (-\t){\underline c}_{(-)}
- {\rm sign}(\t)\, 2a\,{\underline y}_0 \Bigl(\g_0 - \h + \h\ln a|\t|\Bigr)\Bigr\rb \; ,
\lab{y-0} \\
{\underline \ydot}(\t) \approx \th (\t){\underline c}_{(+)} - \th (-\t){\underline c}_{(-)}
- {\rm sign}(\t)\, 2a\, {\underline y}_0 \Bigl(\g_0 + \ln a|\t|\Bigr) \; ,
\lab{y-1} \\
{\underline \ydotdot}(\t) \approx - {\underline y}_0 \frac{a}{|\t|} +
\d (\t) \Bigl\lb {\underline c}_{(+)} + {\underline c}_{(-)} 
-2a \Bigl( 2\g_0 + \ln a + {\widehat c}\Bigr){\underline y}_0 \Bigr\rb \; ,
\lab{y-2}
\er
where ${\widehat c}$ is a regularization (``cut-off'') dependent constant in the 
renormalization of the the singular distribution product 
$''\ln |\t|\,\d(\t)'' = {\widehat c}\,\d (\t)$ .\foot{This is done along the lines of
the configuration space renormalization of ultraviolet-divergent 
Feynman diagrams in standard quantum field theory employing the original
ideas of Stueckelberg-Peterman-Bogoliubov. They formulate renormalization of
products of distributions with coinciding singularities by first defining
them (upon appropriate regularization) on the configuration space with the 
singularity subset removed, and then extending the result to the whole space
in the spirit of the general theory of distributions 
(for recent developments see Ref.\ct{NST-2013} and references therein).}

Substituting \rf{y-0}--\rf{y-2} back into \rf{y-eqs} we obtain
(recall $\r_0 = |{\underline y}_0 |$):
\be
{\underline c}_{(+)} + {\underline c}_{(-)} = {\underline y}_0 
\Bigl\lb 2a \Bigl( 2\g_0 + \ln a + {\widehat c}\Bigr)
-\Bigl(\frac{\cE}{\r_0} \frac{d h}{d\r}\bgv_{\r=\r_0} - 2a\ln\r_0\Bigr)\Bigr\rb \; ,
\lab{refract}
\ee
whereas substituting \rf{y-large-tau} and \rf{y-0}--\rf{y-1} into the expression for 
$\cJ$ \rf{Noether} yields:
\be
\cJ = {\underline\b}_{(\pm)} \times {\underline\a}_{(\pm)} =
\pm {\underline y}_0 \times {\underline c}_{(\pm)} \; .
\lab{angular-mom}
\ee
The compatibility of the last equalities in \rf{angular-mom} follows from \rf{refract}.

In ordinary boosted Schwarzschild case \ct{dray-thooft} the transverse space 
coordinates ${\underline y} (\t)$ experience a refraction (finite
discontinuity of ${\underline \ydot} (\t)$) at the shock 
wave (\textsl{i.e.}, at $\t=0$). In the present case ${\underline y} (\t)$
experience reflection when meeting the shock wave due to the logarithmic singularity
at $\t=0$ of ${\underline \ydot} (\t)$ according to Eq.\rf{y-1}:
\be
{\underline \ydot}(\t) \approx -a\,{\underline y}_0 {\rm sign}(\t) \ln |\t| 
\to \pm \infty \;\; {\rm for} \;\; \t \to \pm 0 \; .
\lab{ydot-singular}
\ee
Since the logarithmic singularity \rf{ydot-singular} is an integrable one, 
inserting \rf{y-1} into Eq.\rf{v-integr} we find the same type 
of finite discontinuity in $v(\t)$ at the shock wave 
as in the ordinary boosted Schwarzschild or Reissner-Nordstr{\"o}m case 
\ct{dray-thooft,lousto-sanchez}:
\br
v(\t) \approx  v_0 - h(\r_0)\th (\t) + 
\frac{a^2 \r_0^2}{2\cE}\,\t\,\bigl(\ln|\t|\bigr)^2 \quad 
{\rm for ~small}\;\; \t \; ,\; v_0 = {\rm const} \; ,
\lab{v-small-tau} \\
\llb v\rrb_0 \equiv
\lim_{\epsilon \to 0} \bigl( v(\epsilon) - v(-\epsilon)\bigr) = - h (\r_0) \; .
\lab{v-jump}
\er
Using Eq.\rf{y-large-tau} implies for large $\t$:
\be
v(\t) \approx \left\{\begin{array}{ll} 
{\frac{1}{2a\cE}\,\bigl(m_0^2 +{\underline\b}^2_{(+)}\bigr)\, e^{a\t}
\to +\infty \;\; ,\;\; {\rm for}\; \t \to +\infty}
\\ \phantom{aaa} \\
- {\frac{1}{2a\cE}\,\bigl(m_0^2 +{\underline\b}^2_{(-)}\bigr)\, e^{a|\t|} 
\to -\infty \;\; ,\;\; {\rm for}\; \t \to -\infty}
\end{array}\right.
\; .
\lab{v-asymp}
\ee

Taking into account Eqs.\rf{u-sol}, \rf{u-confined}, \rf{y-large-tau}, \rf{y-0}, 
\rf{v-small-tau}, \rf{v-asymp}
describing the test particle trajectory, we conclude that the
gravitational electrovacuum shock wave \rf{UR-RN-metric}--\rf{UR-EM-field} 
{\em confines}~ charged particles 
with charges $q_0$ opposite to the pre-boost Reissner-Nordstr{\"o}m charge $Q$.
Namely, according to \rf{u-confined} 
$|u(\t)|\leq \frac{2\cE}{a} \;\; {\rm for ~all}\; \t \in (-\infty,+\infty)$, 
and thus we deduce the size of the confining distance to be equal to
$2\cE/a \equiv 2\sqrt{2} \cE/q_0 f_0 $ for both massive and massless charged test 
particles.


\section{Conclusion}
\label{conclude}

In the present note we studied the result of applying the Aichelburg-Sexl 
ultra-relativistic boost procedure to the static spherically symmetric solutions 
of a non-standard Reissner-Nordstr{\"o}m type obtained from gravity coupled to a 
special kind of nonlinear gauge field system containing a ``square-root'' Maxwell
term \rf{gravity+GG}, which is known to produce charge confinement in flat
space-time. 

Unlike the case of the ultra-relativistic boost of ordinary Reissner-Nordstr{\"o}m
black hole solution \ct{lousto-sanchez}, where the electromagnetic field
disappears in the boost limit leaving only a residual term in the
energy-momentum tensor, in the present case due to the
``confinement''-inducing ``square-root'' Maxwell term the electromagnetic
field persists in the ultra-relativistic limit as a non-trivial ``vacuum''
solution \rf{E-B-shock} of the nonlinear electromagnetic field equations of
motion -- a solution, which does not exist in ordinary Maxwell
electrodynamics. The resulting ``boosted'' solution describes a
gravitational electrovacuum shock wave which possesses the remarkable
property of confining (trapping) charged test particles (both massive and
massless) with charges opposite to the pre-boost-limit black hole charge.
 
The above obtained gravitational electrovacuum shock wave plus the trapped 
charged particles could give us a picture of a hadron in the 
ultra-relativistic limit and one interesting generalization of these
calculations could be solving for a test Dirac particle instead of for a
spinless point particle in the background of such electrovac-gravity shock wave.

Another possible extension of our results would be the study of collisions of two 
electrovac-gravity shock waves.

Finally, a very interesting issue that deserves further attention concerns the 
possible processes of pair creation and of pair annihilation in the presence of 
the gravitational electrovacuum shock wave solution. 

\appendix
\section{}

Unlike confinement of charged particles (massive and massless ones)
when initially they are close enough to the shock wave ($|u_0|< 2\cE/a$
implying $|u(\t)|< 2\cE/a$ for all $\t$), the initial condition $|u_0|> 2\cE/a$
produces the following solution of Eq.\rf{u-eq} (here we take for definiteness
positive $u_0$, the trajectory for negative $u_0$ is obtained by changing
$(u,v) \to (-u,-v)$):
\br
u(\t) = \frac{2\cE}{a} + \Bigl( u_0 - \frac{2\cE}{a}\Bigr) e^{-a\t} \; ,
\lab{u-sol-AA} \\
u(\t) > \frac{2\cE}{a} \;\; {\rm for ~all}\;\t \quad ,\quad
u(\t) \to \left\{\begin{array}{ll} {+\infty \;\; ,\; \t \to -\infty} \\
{2\cE/a \;\; ,\; \t \to +\infty}
\end{array}\right.
\lab{u-asymp-AA}
\er
In particular, the trajectory never comes close to the shock wave at $u=0$,
\textsl{i.e.}, no delta-function terms appear in \rf{y-eqs} and \rf{v-eq}, therefore,
there are no ``refraction'' of ${\underline y}(\t)$ and no discontinuity in
$v(\t)$ at $\t=0$. The large $|\t|$-asymptotics of $v(\t)$ is:
\be
v(\t) \to
\left\{\begin{array}{ll}
{+\infty \;\; {\rm when} \; \cJ \neq 0} \\
{v_0 = {\rm const} \;\; {\rm when} \; \cJ = 0}
\end{array}\right. \;\; {\rm for}\; \t \to - \infty \; ,
\lab{v-asymp-AA-minus}
\ee
\be
v(\t) \to - \infty \;\; {\rm for}\; \t \to + \infty \; ,
\lab{v-asymp-AA-plus}
\ee
where $\cJ$ indicates the conserved angular momentum. The last relations
show that the proper time (world-line parameter) $\t$ flows opposite to the
``laboratory'' time $t$, therefore, according to \ct{QED-feynman} we have to
interpret the resulting trajectory given by \rf{u-sol-AA}-\rf{v-asymp-AA-plus} 
as a motion of an anti-particle which is completely repelled by the shock wave. 

\section*{Acknowledgments}
We gratefully acknowledge support of our collaboration through the academic exchange 
agreement between the Ben-Gurion University and the Bulgarian Academy of Sciences.
S.P. has received partial support from COST action MP-1210.


\end{document}